\documentstyle{article}
\everymath{\rm }
\everydisplay{\rm }

\begin{document}
\begin{center}
{\LARGE Closure of Constraints for \\ [.125in] Plane
Gravity Waves} \\ [.25in]
\large Donald E. Neville \footnote{\large e-mail address:
nev@vm.temple.edu }\\Department of Physics \\Temple University
\\Philadelphia 19122, Pa. \\ [.25in]
July 1, 1995\\ [.5in]
\end{center}
\newcommand{\E}[2]{\mbox{$\tilde{{\rm E}} ^{#1}_{#2}$}}
\newcommand{\A}[2]{\mbox{${\rm A}^{#1}_{#2}$}}
\newcommand{\N}{\rm N}
\newcommand{\Np}{\mbox{${\rm N}'$}}
\newcommand{\M}{\rm M}
\newcommand{\Etwo}{\mbox{$^{(2)}\!\tilde{\rm E} $}}
\newcommand{\Etld }{\mbox{$\tilde{\rm E}  $}}
\def \ut#1{\rlap{\lower1ex\hbox{$\sim$}}#1{}}
\newcommand{\phst}{\mbox{$\phi\!*$}}
\Roman{section}
\large
\begin{center}
{\bf Abstract}
\end{center}
     The metric for gravitational plane waves has very high
symmetry (two spacelike commuting Killing vectors).  For this
high
symmetry, a simple renormalization of the lapse function is found
which allows the constraint algebra for canonical quantum gravity
to close; also, the vector constraint has the correct form to
generate spatial diffeomorphisms.  A measure is constructed which
respects the reality conditions, but does not yet respect the
invariances of the theory.
\clearpage

\section{Introduction}
     The connection-triad variables introduced by Ashtekar
\cite{Ash87} have
simplified the constraint equations of quantum gravity; further,
these variables suggest that in the future we may be able to
reformulate gravity in terms of non-local holonomies rather than
local field operators \cite{RovSmo, GambTri}.  However, the new
variables are
unfamiliar, and it is not always clear what they mean physically
and geometrically.  For example, it is not clear what operators
or structures correspond to gravity waves.  On the other hand,
there has been great progress in constructing diffeomorphism-
invariant volume and area operators; these operators turn out to
be essentially counting operators for numbers of loops and loop
intersections \cite{AV}.  The quantum constraint equations are
much
simpler
in the new variables, and solutions to these equations have been
found \cite{RovSmo, knotsol} .  However, these solutions
correspond
to
some metric, and it is not always clear what that metric is (the
problem of physical interpretation again).  In order to
investigate the metric, one needs a measure, so as to be able to
form dot products and take expectation values.  Some progress has
been made in constructing a measure \cite{measure}.

     The new approach is not yet truly non-local, and it shares
the same renormalization and regularization difficulties which
plague other local field theories such as QCD.  For gravity,
there are some new twists to the old story.  The theory is hard
to regulate, because the regulators do not always respect
diffeomorphism invariance \cite{Blen, OptrOrd, Bor}.   However,
the
theory is astoundingly
easy to renormalize \cite{RovSmo}.  (Compare to QCD, which is
easy
to regulate,
whereas renormalization is difficult.)

     Also, it is not always possible to operator-order the
gravitational constraints so that both the constraint algebra
closes (commutator of two constraints = sum of constraints) and
the vector constraints generate spatial diffeomorphisms
\cite{knotsol}.  (This
is a difficulty which the Ashtekar approach shares with more
traditional approaches.)  The constraint closure is not only
essential to the consistency of the Dirac quantization procedure;
closure is important even classically.  When the 3 + 1 splitup is
integrated forward in time, to construct the entire spacetime,
the theory will not be invariant under the full four-dimensional
diffeomorphism group unless the constraint algebra closes
\cite{HKTei}.  Thus
an invariance of the classical theory is lost if commutator
closure is neglected.

     In this paper we consider the application of the Ashtekar
formalism to the problem of plane gravitational waves.  "Plane"
means the metric possesses two commuting spacelike Killing
vectors, and we shall choose coordinates so that these vectors
are unit vectors pointing in the x and y directions.

\begin{equation}
     k^{(x)} = \partial _x; k^{(y)} = \partial _y.
\label{1.0}
\end{equation}

     We begin the quantization procedure in section 2 by
choosing a factor ordering and verifying closure of the
constraint algebra.  We find a rather surprising result: in the
plane wave case, one and the same factor ordering makes the
vector constraint into a diffeomorphism generator and allows the
algebra to close.  This result requires the high symmetry (the
two spatial Killing vectors).

     In section 3 we tackle the problem of constructing a
measure.  Here we attain some partial success.  Our measure
respects the reality constraints obeyed by the Ashtekar
connections.  However, our measure is not invariant under the
constraints.

     The literature on plane waves is vast, but we single out two
papers which are especially close to the present paper.  Husain
and Smolin \cite{HSm} were the first to apply the Ashtekar
formalism to the two-Killing vector case.
Neville \cite{Nev}, working with the traditional geometrodynamic
variables,
found the transformations which reduce the Hamiltonian to
parameterized free field form and constructed the classical
constants of the motion.  This paper will be referred to as I.

     Reference I studied waves which were unidirectional as well
as plane, where "unidirectional" in the present coordinates means
the waves are moving only in the +z direction.  Since
unidirectional waves are known to obey a superposition principle,
they do not scatter, except off waves moving in the -z direction.
The present paper does not use the  unidirectional assumption.
All results on closure and the measure hold in the presence of
scattering.

     Our notation is typical of papers based upon the Hamiltonian
approach with concomitant 3 + 1 splitup.  Upper case indices A,
B, $\ldots $,I, J, K, $\ldots$ denote local Lorentz indices
("internal" SU(2) indices) ranging over X, Y, Z only.  Lower case
indices a, b, $\ldots $, i, j, $\ldots $ are also three-
dimensional and denote global coordinates on the three-manifold.
Occasionally the formula will contain a field with a superscript
(4), in which case the local Lorentz indices range over X, Y, Z,
T and the global indices are similarly four-dimensional; or a
(2), in which case the local indices range over X, Y (and global
indices over x, y) only.  The (2) and (4) are also used in
conjunction with determinants; e.\ g., g is the usual 3x3
spatial determinant, while\ $^{(2)}e$ denotes the determinant of
the 2x2 X, Y subblock of the triad matrix $e^A_a$. We use Levi-
Civita symbols of various dimensions: $\epsilon _{TXYZ} =
\epsilon _{XYZ} = \epsilon _{XY} = +1$.  The basic variables of
the Ashtekar approach are an inverse densitized triad \E{a}{A}
and a complex SU(2) connection \A{A}{a}.

\begin {eqnarray}
     \E{a}{A}& =& \sqrt g e^a_A; \\
\label{eq:1.1}
     [\E{a}{A},\A{B}{b}]&=& -\hbar \delta (x-x') \delta ^B_A
\delta ^a_b.
\label{eq:1.2}
\end{eqnarray}

The local Lorentz indices are vector rather than spinor; strictly
speaking the internal symmetry is O(3) rather than SU(2), gauge-
fixed to O(2) rather than U(1).

\section{Closure of Constraints}

     Since the result that the constraints do close is
surprising, and since any proof of closure is bound to be
detailed, it would be helpful if we could state some simple
reason why the constraints close, before becoming enmeshed in the
details.  This is not too hard to do.   We consider both the
geometrodynamical and Ashtekar approaches, since closure should
be
independent of choice of basis variables.

     We begin by establishing some notation.  In the coordinate
system (1.1), metric components are independent of x and y, and
it is possible to bring the metric to a block diagonal form
\cite{EhlK}.  One
2x2 subblock connects only t and z components of the metric;
another 2x2 subbolck connects only x and y components.  The x,y
subblock
may be parameterized using variables suggested by
Szekeres \cite{Sz}:

\[
     ds^2 = e^A [dx^2 e^B  \cosh{W} + dy^2 e^{-B} \cosh{W} -
2dxdy
\sinh{W} ]
\]

\begin {equation}
+ e^{D-A/2} \{ [ -{(\Np)}^{2}  + {(\N ^z)}^{2}]dt^2 +2 \N^z dzdt
+
dz^2 \}.
\label{eq:2.1}
\end{equation}
\Np\ is not quite the usual lapse N.

\begin{equation}
     \Np = \N /\sqrt{g_{zz}} .
\label{eq:2.2}
\end{equation}

     Now recall the usual reason why the constraints do not
close.  Let the Hamiltonian density be $\N H_0 + \N ^i H_i$, with
$H_0$ the scalar constraint and $H_i$ the vector constraints.
Then the commutator of two scalar constraints,

\begin{equation}
     [\N H_0,\M H_0]  =  \delta(x - x')[\M\partial_i\N  - \N
\partial_i\M ]H^j g^{ij},
\label{eq:2.3}
\end{equation}
does not close because the final inverse metric factor prevents
$H_j$ from annihilating the wave funtional, implicitly assumed to
stand to
 the right on both sides of equation~(\ref{eq:2.3}).  In the
present case,
from equation~(\ref{eq:2.1}), the inverse spatial metric is
block diagonal, with a 1x1 subblock containing only
$g^{zz}$.  Also, x and y vector constraints have been gauged away
in the process of bringing the metric to block diagonal form, so
that the problem contains only a z vector constraint, and the
$g^{ij}$ in equation ~\ref{eq:2.3} collapses to $g^{zz} =
1/g_{zz}$.  Thus we can completely eliminate the $g^{ij}$ problem
by absorbing a factor of $\sqrt{1/g_{zz}} $ into each of the
lapse functions N and M, as at equation~(\ref{eq:2.2}).  This is
the simple
reason why the constraints close.  The innocuous-looking
renormalization~(\ref{eq:2.2}) is the key step.

     What would the renormalization~(\ref{eq:2.2}) look like in
the Ashtekar notation?  In that formalism the 3-metric and its
conjugate momenta are replaced by complex connection fields
\A{A}{a} and densitized inverse triads \E{a}{A}, where A = X,Y,Z
is a local Lorentz index.  Besides the vector and scalar
constraints, there are three new constraints, which generate
local SU(2) rotations (or in our case, local O(3) rotations,
since A is a vector rather than a spinor index.),  The 3x3
\E{a}{A}
matrix may be brought to block diagonal form, with one 1x1
subblock
plus one 2x2
subblock, exactly as for the 3x3 spatial subblock of the metric
$g^{ij}$  The \E{a}{A} matrix is not symmetric, however, so
contains
more independent components than the metric.  To bring the
inverse
triad
matrix to block diagonal form , one must gauge away $9 - (1 + 4)
= 4$ inverse triad components, whereas in the metric case one
removes
only $6 - (1 + 3) = 2$ components.  To remove the extra two
components one must fix (the x and y vector constraints, as
before, plus) two more constraints, the X and Y SU(2) or O(3)
generators.  The 1x1 subblock of the \E{a}{A} matrix is occupied
by
\E{z}{Z} , while the 2x2 subblock contains all \E{a}{A} with a =
x,y and
A = X,Y.  The Ashtekar scalar constraint H is density weight 2,
and the new Lagrange multiplier for
the scalar constraint is a weight -1 lapse \ut{N} which is
related
to \Np as follows:

\begin{eqnarray}
     \ut{N} \E{z}{Z} &=& (\N /\sqrt{g})
                    (\sqrt{g} e^z_Z) \nonumber \\
                    &=&  \N \sqrt{g^{zz}} \nonumber \\
                    &=& \Np .
\label{eq:2.4}
\end{eqnarray}
 Hence in the Ashtekar approach we will be absorbing factors of
\E{z}{Z} into the densitized lapse.

     Since $\ut{N} H = \Np (H/\E{z}{Z}) $, the new scalar
constraint
$H/\E{z}{Z} $ becomes a rational function of the basic fields,
rather than a polynomial function.  This complication is a price
which must be paid in order to secure closure.  We consider it a
small
price.  Firstly, the constraints should close as a matter of
principle, in order to have a consistent quantization and full
four-dimensional diffeomorphism invariance.

     Secondly, the presence of the $1/\E{z}{Z} $ factor actually
makes the Hamiltonian \underline{less} singular.  $H$ contains
products of several operators, all evaluated at the same point z,
which can lead to undefined $\delta (z-z)$ factors when $H$ acts
upon a wave functional.  To avoid such factors, Husain and Smolin
\cite{HSm}
must regulate $H$ by point-splitting.  However, most of the terms
in $H$ contain a factor of $\E{z}{Z} $, which is removed by the
$1/\E{z}{Z} $, leaving behind a simpler operator which cannot
produce a $\delta (z-z)$ and does not have to be regulated.  To
study the term where the $1/\E{z}{Z} $ does not cancel, we recall
that \E{z}{Z} is conjugate to the complex Ashtekar connection
\A{A}{a} ,

\begin{equation}
     [\E{a}{A} ,\A{B}{b} ] = -\hbar \delta (z - z')\delta ^a_b
\delta ^B_A,
\label{eq:2.7}
\end{equation}
 Therefore the action of \E{z}{Z} on a wave functional $\psi =
\psi [{\rm A}]$ is $\E{z}{Z} \psi = -\hbar \delta \psi / \delta
\A{Z}{z} $. We expect the \A{Z}{z} dependence of $\psi $ to be
holonomic.

\begin{eqnarray}
     \psi & =& \cdots X(z_3) \exp [i\int_{z2}^{z3} \A{Z}{z} S_Z
dz]X(z_2)\exp [i\int_{z1}^{z2} \A{Z}{z} S_Z dz] \cdots \nonumber
\\
     &:=& \cdots X(z_3)U(z_3, z_2)X(z_2)U(z_2,z_1)\cdots .
\label{eq:2.8}
\end{eqnarray}
The X are assumed to be operators in the Lie algebra of SU(2),
and independent of \A{Z}{z} .  Explicit factors of i are present,
because we use the usual, Hermitian SU(2) generators $S_I$.  Then
the direct action of  \E{z}{Z} on the wave functional is

\begin{equation}
\E{z}{Z} (z)\psi = \cdots + \cdots X(z_3)[\hbar \theta (z_3-z)
\theta (z-z_2) U(z_3, z_2)(-iS_Z)]X(z_2)U\cdots + \cdots ,
\label{eq:2.8a}
\end{equation}
one such term for each U.  To  find the action of
$[1/\E{z}{Z}]$, multiply both sides of equation~(\ref{eq:2.8a})
by $\theta (z_3-z)\theta (z-z_2)$, in order to project out the
term
exhibited
explicitly on the right; then multiply both sides of
equation~(\ref{eq:2.8a})
by $\hbar /\E{z}{Z}$.

\begin{eqnarray}
[\hbar /\E{z}{Z} (z)] &\cdot &\theta (z_3-z)\theta (z-z_2)[\cdots
X(z_3)U(z_3, z_2)(-iS_Z)X(z_2)
     \cdots ] \nonumber \\
& = & \theta (z_3-z)\theta (z-z_2)[\cdots X(z_3)U(z_3,
z_2)X(z_2)\cdots ] + \nonumber \\
 & & + const.
\label{eq:2.8c}
\end{eqnarray}
In order to make this look like $\hbar /\E{z}{Z}$ acting upon
$\psi$,
multiply $\psi$ by the following partition of unity:
\begin{equation}
1 = [\theta (z-z_n) + \theta (z_n-z)\theta(z-z_{n-1}) + \cdots +
      \theta (z_3-z)\theta (z-z_2) + \cdots ].
\label{eq:2.8d}
\end{equation}
Therefore
\begin{eqnarray}
[\hbar /\E{z}{Z} (z)]\psi  & = & \cdots + \cdots X(z_3)[\theta
(z_3-
z) \theta (z-z_2) U(z_3, z_2)(-iS_Z)^{-1}]X(z_2)U\cdots  +
\nonumber
\\
& & +  \cdots  + {\rm const.}.
\label{eq:2.8b}
\end{eqnarray}
Evidently the action of $\hbar /\E{z}{Z}$ on $\psi$ is quite
mild.
There
is not even a $\delta (z-z')$ factor, let alone a $\delta (z -
z)$
{}.

     If the X are helicity-changing operators,
the eigenvalue of $S_Z$ in equation~(\ref{eq:2.8b}) will vary
from one U to the next.  If we use the 2x2 Pauli representation,
the eigenvalue never vanishes and $(S_Z)^{-1}$ is always
finite.  However, in future work we shall use the (2j+1)x(2j+1)
representation, where $S_Z$ can have a zero
eigenvalue if j is integer.  When $S_Z$ vanishes, so that
$U(z_3,z_2)$ is unity, one may
replace the square bracket in equation~(\ref{eq:2.8b})  by
\begin{equation}
[-\theta (z_3-z) \theta (z-z_2) U(z_3, z_2)\int_2^3 \A{Z}{z} dz].
\label{eq:2.8e}
\end{equation}
As a check, when equation~(\ref{eq:2.8b}) is inverted by
multiplying both sides
by  $\E{z}{Z}/\hbar $, the square bracket~(\ref{eq:2.8e}) gives
the
same answer
for vanishing $S_Z$ as the square bracket~(\ref{eq:2.8b}) gives
for
finite  $S_Z$.

     We should also check that \E{z}{Z} is not a gauge artifact
(which can be gauged to zero!).  Despite its contravariant z
index, \E{z}{Z} is a scalar function under diffeomorphisms,

\begin{equation}
     \delta \E{z}{Z} = \N ^z \partial _z \E{z}{Z}
\label{eq:2.5}
\end{equation}
The \E{z}{Z} field is both contravariant and weight 1, and in
effect the two transformation properties cancel each other in one
space dimension, leaving an ordinary scalar.  Another way to see
the scalar
behavior is to relate \E{z}{Z} to the metric variables A, B, W,
and D introduced at equation~(\ref{eq:2.1}).  Using the same
relationships
as at equation~(\ref{eq:2.4}), we find

\begin{equation}
     \E{z}{Z} = exp(A).
\label{eq:2.6}
\end{equation}
Since A occurs in the x,y sector of the metric~(\ref{eq:2.1}),
and
this sector transforms as a scalar under diffeomorphisms, the
function A is a scalar.  Hence $\E{z}{Z} $ cannot be gauged away.

     We now pass to the details.  In a coordinate system where
both \E{a}{A}  and \A{A}{a} fields are block diagonal, the total
Hamiltonian reduces to

\begin{eqnarray}
     H_T &=& \Np [\epsilon _{MN}\E{x}{M} \E{y}{N} (\E{z}{Z}
)^{-1}
\epsilon _{AB} \A{A}{x} \A{B}{y} + \epsilon _{MN} \E{b}{M} {\rm
F}^N_{zb}]
\nonumber \\ & & + i\N ^z\E{b}{M} {\rm F}^M_{zb} \nonumber \\
     & & -i{\rm N_G} [\partial _z\E{z}{Z} - \epsilon
_{IJ}\E{a}{I} \A{J}{a} ] \nonumber \\
     &\equiv & \Np H_S + \N^zH_z + {\rm N_G}H_G,
\label{eq:2.9}
\end{eqnarray}
where\begin{equation}
     {\rm F}^N_{zb} = \partial _z\A{N}{b}
                          -\epsilon _{NQ}\A{Z}{z} \A{Q}{b},
\label{eq:2.10}
\end{equation}
and the ${\rm N_G}$ term (the Gauss constraint) is the generator
of local SU(2) rotations around the Z axis.  The lapse has been
renormalized as at equation~(\ref{eq:2.4}), by removing a factor
\E{z}{Z}
from the scalar constraint.  From now on the "scalar constraint"
will mean the expression $H_S$ multiplying \Np\ in
equation~(\ref{eq:2.9}), namely, the usual Ashtekar scalar
constraint H
divided by \E{z}{Z} .  We have operator-ordered
equation~(\ref{eq:2.9}) in
a way which anticipates the following section, where we shall
consider solutions $\psi $ which depend on \A{Z}{z} and the four
\Etld\ in the 2x2 sector.  If we call these five commuting
variables
the Q variables ($\psi = \psi (Q)$) and the five conjugate
variables the P variables, then we have ordered P's to the right,
Q's to the left in $H$.  We shall carry out the proof of closure
for this specific choice of Q's and this specific ordering, but
the proof would also go through for the other popular choice of
Q's. in which the five A's are chosen as Q's (and the A's are
ordered to the left in $H$).

     Now let us ask which commutators, or which parts of which
commutators, are likely to give trouble.  First of all, it is
easy to check that the \underline{classical} commutators (or
rather, the classical Poisson brackets) close on pure
constraints, with no undesirable factors of $g^{zz}$ or the
Ashtekar analog of $g^{zz}$.  Since the quantum commutators are
designed to reproduce exactly the same fields as the classical
Poisson brackets, there will be no factor of $g^{zz}$ in the
quantum case either.  We will get the same fields; and the only
thing which can go wrong is that the commutator yields a P field
to the left of a Q field.  Remembering that each constraint is a
sum of terms of the form f(Q)g(P), we want, schematically,

\begin{equation}
     [f_1(Q)g_1(P),f_2(Q)g_2(P)] = f_3(Q)g_3(P).
\label{eq:2.11}
\end{equation}
There would be trouble if P's occured to the left of Q's on the
right-hand side, for example $Pf_3(Q)g_3(P)$.

     We now show that almost all the terms in a typical
constraint commutator $[H_i,H_j]$ will give no trouble.  Each
term in this commutator will look like the left-hand side of
equation~(\ref{eq:2.11}).
Write this term out using the identity

\begin{eqnarray} \label{eq:2.12}
     [AB,CD]& =& AC[B,D]+A[B,C]D+ \nonumber \\
               & &  +C[A,D]B+[A,C]DB. \\
\label{eq:2.13}
     [f_1g_1,f_2g_2]&=& 0 + f_1[g_1,f_2]g_2 + f_2[f_1,g_2]g_1 +
0.
\end{eqnarray}
On the right-hand side, the f and g factors outside the
commutators

are in the correct order (Q's to the left).  Therefore if
the commutators on the right yield fields which commute among
themselves, the entire expression on the right can be ordered
correctly, and the term gives no trouble.  In particular, if
either f or g is a monomial, say $g\sim $P, then the commutator
of any f with g yields only Q fields, and the term can be ordered
correctly.  Examination of equation~(\ref{eq:2.9}) shows that all
the terms in $H_T$, except the $(\E{z}{Z} )^{-1}$ term, are
monomials in the P's, of the form $Q^2P$ or $QP$; and one term is
independent of the P's (pure Q) which is even better.  Therefore
the only commutators we have to check are the ones involving the
$(\E{z}{Z} )^{-1}$ term,

\begin{eqnarray}
     H_E &:= &\Np [\epsilon _{MN}\E{x}{M} \E{y}{N} (\E{z}{Z} )^{-
1}\epsilon _{AB} \A{A}{x} \A{B}{y} \nonumber \\
     &\sim &Q^2(1/P)P^2.
\label{eq:2.14}
\end{eqnarray}

     We now investigate commutators of $H_E$ with Q terms,
QP terms, $Q^2P $ terms, and finally commutators of
$H_E$ with itself.  (a). Commutators of $H_E$ with Q and
QP terms.  For this case, in commutator~(\ref{eq:2.11}),
\underline{both} $f_2$ and $g_2$ are monomials or constants.  It
follows immediately that both commutators on the right-hand side
of equation~\ref{eq:2.13} involve at least one monomial and can
be
correctly ordered.  Even if $f_2$ is \A{Z}{z} , there is no
problem, since the commutator with (\E{z}{Z} )$^{-1}$,

\begin{equation}
 [(\E{z}{Z} )^{-1},\A{Z}{z} ] = \hbar \delta (z-z') (\E{z}{Z}
)^{-2},
\label{eq:2.14a}
\end{equation}
yields factors which commute among themselves.
Equation~(\ref{eq:2.14a})
may be proven by multiplying it on left and right
by \E{z}{Z} .

     (b). Commutators of $H_E$ with $Q^2P$ terms.  These
are terms of the form \E{b}{M} \A{Z}{z} \A{Q}{b} coming from the
F's
(field strengths) in the scalar and vector constraints, equations
{}~(\ref{eq:2.9})-(\ref{eq:2.10}).  (b1).  When commuting the
scalar
constraint with itself, for example, we get commutators of the
form

\begin{eqnarray*}
[Q_E^2(1/P_E)P_E^2(z),Q^2P(z')] &+&
[Q^2P(z),Q_E^2(1/P_E)P_E^2(z')] \\
     &=&\cdots + Q_E^2[(1/P_E)P_E^2(z),Q^2(z')]P
+ \\ & & +Q_E^2[Q^2(z),(1/P_E)P_E^2(z')]P,
\end{eqnarray*}
where subscripts E denote fields coming from $H_E$, and $\cdots $
indicates harmless commutators involving the monomial P.  The two
commutators on the last line differ by a minus sign and an
interchange of z and $z'$.  The interchange affects nothing,
since
the commutator is proportional to $\delta (z-z')$.  Therefore the
two commutators cancel each other, and the term is harmless.
     (b2).  When  commuting the scalar constraint with the vector
constraint, we get

\begin{eqnarray*}
[Q^2P(z),Q^2(1/P)P^2(z')]&=& \cdots +
Q_2(z')[Q^2(z),(1/P)P^2(z')]P(z) \\
     &=& \cdots + \Etwo [\E{b}{M} \A{Z}{z} ,(\E{z}{Z} )^{-
1}\epsilon _{AB} \A{A}{x} \A{B}{y}]i\A{N}{b} \epsilon _{MN} \\
     &=& \cdots + \Etwo \E{b}{M} [\A{Z}{z} ,(\E{z}{Z} )^{-
1}]\epsilon _{AB} \A{A}{x} \A{B}{y} i\A{N}{b} \epsilon _{MN} + \\
& &
 + \Etwo (\E{z}{Z} )^{-1}[\E{b}{M} ,\epsilon _{AB} \A{A}{x}
\A{B}{y}]\A{Z}{z} i\A{N}{b} \epsilon _{MN} \\
     &=& \cdots - \Etwo \E{b}{M} [\hbar \delta (z-z')(\E{z}{Z}
)^{-2}]\epsilon _{AB} \A{A}{x} \A{B}{y} i\A{N}{b} \epsilon _{MN}
+0. \end{eqnarray*}
On the third line we have used the ABCD identity,
equation~(\ref{eq:2.12}).  On the last line, we can commute the
i\A{N}{b} factor to the left, until it forms the expression
$i\E{b}{M} \A{N}{b} \epsilon _{MN}$.  We can subtract from this
expression the Gauss constraint $H_G$, equation~(\ref{eq:2.9}).
This causes no problems, since the Gauss constraint commutes with
everything to the right of $i\E{b}{M} \A{N}{b} \epsilon _{MN}$,
hence can be commuted to the far right where it will eventually
annihilate the wave functional.  Since $i\E{b}{M} \A{N}{b}
\epsilon _{MN} - H_G = i\partial _z\E{z}{Z} $, the last line
becomes

\begin{displaymath}
\cdots +i\Etwo [\hbar \delta (z-z')\partial _z(\E{z}{Z} )^{-
1}]\epsilon _{AB} \A{A}{x} \A{B}{y}.
\end{displaymath}
This now has the same form as the corresponding term in the
classical calculation, and moreover the operators are correctly
ordered (Q's to the left).

     The calculation just completed, however, suggests a new way
in which the constraints might fail to close.  Suppose that at
some point in the calculation it is necessary to insert the Gauss
constraint in the middle of a term (as was done just above); if
the $H_G$ factor cannot be commuted to the far right, then
closure will be spoiled.  Fortunately, the classical calculation
once again comes to the rescue.  Since the classical calculation
has the same pattern of fields as the quantum calculation, a
Gauss insertion is necessary in the quantum calculation if and
only if a Gauss insertion is necessary at the same point in the
classical calculation.  It turns out that the \underline{only}
point where a Gauss insertion occurs, classically, is in the
[vector,scalar] commutator term just considered, and this
insertion is harmless.

     (c).  Finally, we consider the commutator of $H_E$ with
itself.  Using the ABCD identity~(\ref{eq:2.12}), we get

\[
[H_E(z),H_E(z')] = \Etwo (z)(\E{z}{Z} )^{-1}[\epsilon _{AB}
\A{A}{x} \A{B}{y} (z),\Etwo (z')(\E{z}{Z} )^{-1}]\epsilon _{CD}
\A{C}{x} \A{D}{y} (z')+
\]
\[
\Etwo (z')(\E{z}{Z} )^{-1}[\Etwo
(z)(\E{z}{Z} )^{-1},\epsilon _{CD} \A{C}{x} \A{D}{y}
(z')]\epsilon _{AB} \A{A}{x} \A{B}{y} (z).
\]
The two commutators cancel, after a relabeling AB$\leftrightarrow
$CD in the second commutator.  This completes the proof that the
constraints close.

     It is also easily  verified that the constraint $H_z$
generates
diffeomorphisms in the z direction.  (In fact $H_z$ fails to
generate diffeomorphisms
only when the P's  are ordered to the left \cite{knotsol}.)
$H_z$
is not quite the
diffeomorphism generator, but differs from that generator by a
term
of the
form $N^z\A{Z}{z} H_G$, $H_G$ the Gauss constraint.  A term of
this
form can
be added to $H_z$ without affecting closure, since the added term
is linear
in P and therefore harmless.

     Since there is no factor of $g^{zz}$ or other fields in the
constraint algebra, it is a true Lie algebra (structure
``constants'' are at most $\delta $ functions, not functions of
fields).  The structure of this Lie algebra is very
simple.  It breaks up into two commuting subalgebras generated by
$(H_z \pm H_S)/2$,

\begin{equation}
\int dz\int dz'[M(z)(H_z \pm H_S)/2,N(z')(H_z \pm H_S)/2] \\ =
i\hbar \int dz(M\partial _zN -N\partial _zM)(H_z \pm H_S)/2.
\label{eq:2.15}
\end{equation}
Presumably these generators may be interpreted physically as
displacements along the light cone, in directions (z $\pm $ct).

\section{The Reality Constraints}

     Since the Ashtekar connections are complex, they obey
reality constraints of the form A + A* = 2 Re A = known function
of the \Etld.  For completeness, and to establish certain
detailed formulas which we will need later, we sketch a
derivation of these constraints.  The derivation ends at
equation~(\ref{eq:5.12}).   At equation~(\ref{eq:5.13}) we propose
a measure
to enforce these constraints.

     We start from the four-dimensional connection

\begin{equation}
     2{\rm G}\, ^{(4)}\!\A{IJ}{a}  = \omega ^{IJ}_a +i\epsilon
^{IJ}_{MN}\omega ^{MN}_a,
\label{eq:5.1}
\end{equation}
where G is the Newtonian constant and $\omega $ is the SL(2,C)
Lorentz connection.  After the 3 + 1 splitup \cite{Alect, Rovlect},
one obtains the SU(2)
connection which is canonically conjugate to $\E{a}{S}$
(equation~(\ref{eq:1.2})).

\begin{eqnarray}
     2{\rm G}\A{S}{a} &\:= & \epsilon _{MSN}\, ^{(4)}\!\A{MN}{a}
\nonumber \\
     &=&\epsilon _{MSN}\omega ^{MN}_a - 2i\omega ^{TS}_a.
\label{eq:5.2}
\end{eqnarray}
{}From this, the real part of A is

\begin{equation}
     {\rm G}[\A{S}{a} + \A{S}{a} *] = \epsilon _{MSN}\omega
^{MN}_a ,
\label{eq:5.3}
\end{equation}
or when these equations are written out for the 1x1 and 2x2
subblocks,

\begin{eqnarray}
     {\rm G}[\A{Z}{z} + \A{Z}{z} *] &=&-2\omega ^{XY}_z; \\
\label{eq:5.3a}
     {\rm G}[\A{I}{i} + \A{I}{i} *]&=&2\epsilon _{IJ} \omega
^{ZJ}_i .
\label{eq:5.3b}
 \end{eqnarray}
These ``reality constraints'' relate the A's to the \Etld 's ,
since $\omega = \omega (\tilde {\rm E} )$. The next step is to
exhibit
this $\tilde {\rm E}$ dependence.  This is done by  first relating
the
$\omega$ 's to the triads $e^I_i $ and inverse triads $e^i_I$, then
relating the $ e^I_i$ and $ e^i_I$ to the $ \tilde {\rm E}$ .  The
requirement that
the triads have zero covariant derivative leads to
\begin{equation}
     \omega ^{IJ}_a = [\Omega _{i[ja]} + \Omega _{j[ai]} - \Omega
_{a[ij]}]e^{iI} e^{jJ},
\label{eq:5.4}
\end{equation}
where

\begin{equation}
     \Omega _{i[ja]} = e_{iM}[\partial _je^M_a - \partial
_ae^M_j]/2.
\label{eq:5.5}
\end{equation}
In the present case the triad matrix is block diagonal, with 2x2
and 1x1 subblocks, and these equations simplify considerably.
Also, from equations ~(\ref{eq:5.3a}) and ~(\ref{eq:5.3b}) we shall
need only

\begin{eqnarray}
     \omega^ {XY}_z&=& [e^X_i\partial _ze^{Yi} - e^Y_i\partial
_ze^{Xi}]; \label{eq:5.6} \\
     \omega ^{ZJ}_i &=& -\partial _zg_{ij}e^{zZ}e^{jJ}/2,
\label{eq:5.7}
\end{eqnarray}
where the indices i,j range over x,y only.

     Now we replace metric, triads, and inverse triads by \Etld\
fields.  The inverse triad fields are easiest to replace.  From
equation~(\ref{eq:1.1})

\begin{eqnarray}
     e^a_A &=& \E{a}{A} /\sqrt g \nonumber \\
          &=&\E{a}{A} /\sqrt{\E{z}{Z}\Etwo  }.
\label{eq:5.8}
\end{eqnarray}
For the metric and triad fields, the strategy is to express them
in terms of the inverse triads, then replace the latter.  For
example in the 2x2 subblock,

\begin{eqnarray}
     ^{(2)}g_{ab}& =& \epsilon _{am}\epsilon
_{bn}{}^{(2)}\!g^{mn}\, ^{(2)}\!g\nonumber \\
     &=&\epsilon _{am}\epsilon _{bn} \E{m}{M} \E{n}{M} ^{(2)}\!g/g
\nonumber \\
     &=&\epsilon _{am}\epsilon _{bn} \E{m}{M} \E{n}{M} \E{z}{Z}
/ \Etwo;
\label{eq:5.9}
\end{eqnarray}

\begin{eqnarray}
     e^M_m &=& \epsilon _{MN}\epsilon _{mn}e^n_N\,  ^{(2)}e
\nonumber \\
     &=& \epsilon _{MN}\epsilon _{mn} \E{n}{N} \sqrt{\E{z}{Z}
/\Etwo } .
\label{eq:5.10}
\end{eqnarray}
We make these replacements in equations~(\ref{eq:5.6}) and
{}~(\ref{eq:5.7}) and obtain

\begin{eqnarray}
     \omega^ {XY}_z&=&-[\epsilon _{mn} \E{m}{M} \partial
_z\E{n}{M}]/2 \Etwo ; \\
\label{eq:5.11}
     \omega ^{ZJ}_i &=& -[\E{j}{J} /2\Etwo ]\partial _z[\epsilon
_{im}\epsilon _{jn} \E{m}{M} \E{n}{M} \E{z}{Z} /\Etwo  ].
\label{eq:5.12}
\end{eqnarray}

     After these algebraic preliminaries, we are ready to
consider the measure.  Since our complete set of commuting
observables are the four \Etld\ in the 2x2 X,Y sector, plus the
complex connection \A{Z}{z}, we try a dot product of the form

\begin{equation}
     <\phi \mid \psi > = \int \phst\psi \mu d^4\tilde {\rm E} d^2A,
\label{eq:5.13}
\end{equation}
where $d^2A \equiv dRe\A{Z}{z} dIm\A{Z}{z} $.  The measure $\mu $
must satisfy several requirements.  (i)  It must guarantee the
quantum form of the reality constraints.

\begin{equation}
     <\phi \mid A\psi > + <A\phi\mid \psi > = 2<\phi\mid {\rm
ReA}\psi >.
\label{eq:5.14}
\end{equation}
(ii) It must guarantee the invariance of $\mu d^4\tilde {\rm E}
d^2A$
under transformations generated by the scalar, vector, and Gauss
constraints.  (iii) It must contain enough gauge-fixing delta
functions to remove the usual unbounded integrations over
infinite numbers of gauge copies.  Note that (ii) requires only
invariance under the constraints, not invariance under four-
dimensional diffeomorphisms.  In a 3 + 1 formalism, one does not
have the proper set of fields to implement the latter invariance,
essentially because all fields are evaluated on a constant time
hyperslice, whereas four-dimensional diffeomorphisms move
 fields off the hyperslice \cite{IK}.

     As yet we do not know how to satisfy requirements (ii)-
(iii).  We shall, however, propose a $\mu $ which will satisfy
requirement (i).  We set

\begin{equation}
     \mu = \delta [{\rm G}(\A{Z}{z} + \A{Z}{z}*) + 2\omega
^{XY}_z].
\label{eq:5.15}
\end{equation}
{}From equation~(\ref{eq:5.3a}), this delta function enforces the
\A{Z}{z}
reality constraint.  The surprising fact is that it also enforces
the remaining reality constraints~(\ref{eq:5.3b}) as well, as we
shall prove now.

     First we should clarify our notation.  Since our integration
variables are Re\A{Z}{z} and Im\A{Z}{z} , not A and A*, the
\A{Z}{z} functional derivative really means

\begin{equation}
     \delta /\delta \A{Z}{z} := [\delta /\delta Re\A{Z}{z} +
(1/i)\delta /\delta Im\A{Z}{z} ]/2,
\label{eq:5.16}
\end{equation}
which follows from ReA = (A + A*)/2, etc.  We can then check
that, since the $\phi$ in the bra, equation~(\ref{eq:5.14}) , is
complex
conjugated,

\begin{eqnarray}
     \delta \phst /\delta \A{Z}{z} &=& \delta \phst[{\rm A*}]/
\delta {\rm A} \nonumber \\
     &=&\int dz'[ \delta \phst/\delta \A{Z}{z} (z')\!*][\delta
\A{Z}{z}\!
*/\delta Re\A{Z}{z} + (1/i)\delta \A{Z}{z}\! */\delta Im\A{Z}{z}
]/2 \nonumber \\
     &=& 0,
\label{eq:5.17}
\end{eqnarray}
as expected.  Now write

\begin{eqnarray}
     <\phi \mid \A{I}{i} (z) \psi >& =& -\hbar \int \phst\mu
\delta \psi /\delta \E{i}{I} (z) \nonumber \\
     &=&+\hbar \int [\delta \phst/\delta \E{i}{I} \mu  \psi +
\phst\delta \mu /\delta \E{i}{I} \psi]\nonumber \\
     &=&-<\A{I}{i} \phi \mid  \psi >  + \hbar \int \phst
\delta  \delta [{\rm G(A + A*)} + 2\omega ^{XY}_z]/\delta
\A{Z}{z} (z') \times \nonumber \\
     & & \times [2\delta \omega ^{XY}_z(z')/\delta \E{i}{I} (z)]
dz'\psi \nonumber \\
     &=&-<\A{I}{i} \phi \mid  \psi > - (\hbar /{\rm G})\int \phst
\mu [2\delta \omega ^{XY}_z(z')/\delta \E{i}{I} (z)] dz' \delta
\psi /\delta \A{Z}{z} (z') \nonumber \\
     &=&-<\A{I}{i} \phi \mid  \psi > - (1/{\rm G})\int \phst
\mu [2\delta \omega ^{XY}_z(z')/\delta \E{i}{I} (z)] dz'\E{z}{Z}
(z') \psi .
\label{eq:5.18}
\end{eqnarray}
Again, the $\delta /\delta \A{Z}{z}$ is really a sum of ReA and ImA
functional derivatives, as at equation~(\ref{eq:5.16}), with $
\delta
\mu /\delta Im\A{Z}{z} $ vanishing.  The last square bracket can be
rewritten using equation~(\ref{eq:5.11}) and $^{(2)} \tilde {\rm E}
= \epsilon
_{ij}\epsilon _{IJ}\E{i}{I} \E{j}{J}/2$.

\begin{eqnarray}
-\E{z}{Z} (z')2\delta \omega ^{XY}_z(z')/\delta \E{i}{I} (z) &=&
\E{z}{Z} [(\delta \epsilon _{in}\partial _{z'}\E{n}{I} + \epsilon
_{mi}\E{m}{I} \partial _{z'}\delta )/\Etwo \nonumber \\
     & &  - \epsilon _{mn} \E{m}{M} \partial _{z'}\E{n}{M} \delta
\epsilon _{ij}\epsilon _{IJ}\E{j}{J} /(\Etwo )^2],
\label{eq:5.19}
\end{eqnarray}
where $\delta = \delta (z - z')$.  Now use $\epsilon _{mn}
\epsilon _{ij} = \epsilon _{mi} \epsilon _{nj} +\epsilon _{mj}
\epsilon _{in}$ in the last term of equation~(\ref{eq:5.19}).  The
two terms involving $ \epsilon _{in}$ cancel.  After integrating by
parts to remove the derivative from the delta function in the
second term, one may replace the
$\E{m}{I}$  by $\E{m}{M}\delta^M_I = \E{m}{M} [\epsilon
_{nj}\epsilon
_{IJ}\E{n}{M} \E{j}{J} /\Etwo ]$.  We then have

\begin{eqnarray}
-\E{z}{Z}2 \delta \omega ^{XY}_z(z')/\delta \E{i}{I} &=&-\partial
_{z'}[\E{z}{Z} \epsilon _{mi}\epsilon _{nj}\E{m}{M} \E{n}{M}
/\Etwo  ]\delta (z - z')\epsilon _{IJ}\E{j}{J} /\Etwo \nonumber \\
     &=& 2\omega ^{ZJ}_i \epsilon _{IJ}\delta (z - z') \nonumber
\\
     &=& 2G {\rm Re}\A{I}{i} \delta (z - z'),
\label{eq:5.20}
\end{eqnarray}
where the third line uses equation~(\ref{eq:5.12}) and the last
line uses equation~(\ref{eq:5.3b}).  Inserting the
result~(\ref{eq:5.20}) into equation~(\ref{eq:5.18}), we obtain the
reality condition for the \A{I}{i} field, QED.

\section{Directions for Further Research.}

     For the plane wave problem, we have constructed a constraint
algebra which closes after a simple renormalization of the lapse
function.  We have argued that the cost of this renormalization
(rational, rather than polynomial constraints) is small compared to
the benefits (consistent constraints in the quantum-mechanical
theory; full diffeomorphism invariance in the classical theory).
We have also made modest progress toward constructing a measure.

     It is a standard result that the Hamiltonian has surface terms
whenever the spatial manifold is non-compact \cite{deW}.  In future
work, we intend to describe these surface terms.  They are
surprising: in the plane wave case, it is not automatically true
that the Gauss constraint term in the Hamiltonian falls to zero at
infinity.

     Because of these additional terms at infinity, one must
exercise
care when interpreting any wavefunctional involving
holonomies defined over open contours.  In the planar case a
holonomy $\exp (i\int \A{Z}{z}S_z)$ integrated over a closed
contour is necessarily zero, since the z contour must retrace
itself.  Therefore it is natural to consider open contours and
extend the endpoints to $z = \pm \infty$, to respect spatial
diffeomorphism invariance.  It is possible to enrich this elemental
holonomic structure in two ways.  First, insert \Etld\ operators at
various points along the holonomy; the wavefunctional then looks
like a Rovelli-Smolin $T^n$ operator \cite{RovSmo} (defined over an
open rather than closed contour).  Second, replace the usual 2x2
$S_z$ matrices in the holonomy and in \Etld\ by the (2j + 1) x (2j
+ 1) spin-j representation.   The resultant structure is
reminiscent of a symmetric state, or spin network state
\cite{symmsta} , with the holonomies corresponding to flux lines of
spin j and the \Etld\ operators corresponding to vertices.  For
appropriate choice of the \Etld, a wavefunctional constructed in
this manner is annihilated by the Hamiltonian at {\it finite}
values of z.

     In three spatial dimensions, this would be essentially the end
of the story: the flux exiting at infinity is irrelevant, since
Gauss rotations at infinity are not allowed.  The surviving
surface terms in the Hamiltonian simply give the ADM energy.   In
one spatial dimension, however, the Gauss term contributes at
infinity.  One could add Fermionic matter to the theory
\cite{Romano, Morales} and terminate the flux lines on Fermions at
$\pm \infty$.   However, adding Fermions probably complicates the
theory unnecessarily.  In one-dimensional QED, for example, one can
learn quite a bit by studying electromagnetic plane waves at finite
z, while ignoring what happens to the wave at infinity.  In a
future publication we will adopt this philosophy and study the
finite z properties of the open flux line solutions.  Work is also
in progress on solutions involving closed flux lines.


\begin{thebibliography}{99}
\bibitem{Ash87} Ashtekar A. 1987 Phys. Rev. \underline{D36} 1587
\bibitem{RovSmo} Rovelli C. and Smolin L. 1990 Nucl. Phys.
     \underline{B331} 80
\bibitem{GambTri} for applications of loops to gauge theories in
     general, see Gambini R. and Trias A. 1980 Phys. Rev.
     \underline{D22} 1380; 1986 Nucl. Phys. \underline{B278} 436
\bibitem{AV} Rovelli C. and Smolin L. Discreteness of area and
     volume in quantum gravity.  Report CGPG-94/11/1, gr-qc
     9411005
\bibitem{knotsol} 1992 Br\"{u}gmann B. Gambini R. and Pullin J.
     Phys. Rev. Lett. \underline{68} 431
\bibitem{measure} Ashtekar A. Lewandowski J. Marlof D.
     Mour\~{a}u and Thiemann T. 1994 Quantum geometrodynamics;
     and Coherent state transform of the space of connections.
     Pennsylvania State University preprints.
\bibitem{Blen} Blencowe M. 1990 Nucl. Phys. \underline{B341} 213
\bibitem{OptrOrd} Br\"{u}gmann B. Gambini R. and Pullen J. 1992
     Nucl. Phys. \underline{B385} 587; Br\"{u}gmann B. and
     Pullen J. 1993 Nucl. Phys. \underline{B390} 399
\bibitem{Bor} Borissov R., Regularization of the Hamiltonian
     constraint and closure of the constraint algebra, Report
     TU-94-11, Temple University; gr-qc 9411038
\bibitem{HKTei} Hojman S.A. Kucha\v{r} K. Teitelboim C. 1976 Ann.
     Phys. (NY) \underline{96} 88
\bibitem{symmsta} Rovelli C. Smolin L., Spin networks and quantum
     gravity, Report CGPG-95/1/18.
\bibitem{HSm} Husain V. and Smolin L. 1989 Nucl. Phys.
     \underline{B327} 205
\bibitem{Nev} Neville D.E. 1993 Class. Quantum Grav.
     \underline{10} 2223.  This paper is referred to as I in the
     text.
\bibitem{EhlK} Ehlers J. and Kundt W. 1962 Exact solutions of the
     gravitational field equations, in {\it Gravitation: an
     Introduction to Current Research} edited by Witten L.,
     Wiley, New York
\bibitem{Sz} Szekeres P. 1972 J. Math. Phys. \underline{13} 286
\bibitem{deW} DeWitt B. 1967 Phys. Rev. \underline{160} 1113
\bibitem{Alect} Ashtekar A. (with invited contributions) 1988
     {\it New Perspectives in Canonical Gravity} Bibliopolis
     Naples
\bibitem{Rovlect} Rovelli C. 1991 Class. Quantum Grav.
     \underline{8} 1613
\bibitem{JRS} Jacobson T. and Smolin L. 1988 Nucl. Phys.
     \underline{B299} 295; Renteln P. and Smolin L. 1989 Class.
     Quantum Grav. \underline{6} 275
\bibitem{IK} Isham C.J. and Kucha\v{r} K. 1985 Ann. Phys. (NY)
     \underline{164} 288; Halliwell J.J. and Hartle J.B. 1991
     Phys. Rev. \underline{D43} 1170
\bibitem{Morales} Morales-T\'{e}cotl Hogo A. and Rovelli C. 1994
     Phys. Rev. Letters \underline{72} 3642
\bibitem{Romano} Ashtekar A. Romano J.D. and Tate R.S. 1989 Phys.
     Rev. \underline{D40} 2572
 \end{thebibliography}
\end{document}